\documentclass[%
 reprint,
superscriptaddress,
 amsmath,amssymb,
 aps,
 pre,
]{revtex4-1}

\usepackage{graphicx}
\usepackage{dcolumn}
\usepackage{bm}
\usepackage[colorlinks= true, linkcolor=blue, citecolor=blue, urlcolor=blue]{hyperref}
\usepackage{lipsum}
\usepackage{bbold}
\usepackage{mathtools}
\usepackage[normalem]{ulem}
\usepackage{float}
\usepackage{adjustbox}
\usepackage{cancel}

\def\bea{\begin{eqnarray}}
\def\eea{\end{eqnarray}}

\usepackage{xcolor}

\newcommand{\eref}[1]{Eq.~(\ref{#1})}
\newcommand{\fref}[1]{Fig.~\ref{#1}} 
\begin{document}


\title{Search with stochastic home-returns can expedite classical first passage under resetting}

\author{Arup Biswas}
\email{arupb@imsc.res.in}
\affiliation{The Institute of Mathematical Sciences, CIT Campus, Taramani, Chennai 600113, India}
\affiliation{Homi Bhabha National Institute, Training School Complex, Anushakti Nagar, Mumbai 400094, India}
\author{Anupam Kundu}
\email{anupam.kundu@icts.res.in}
\affiliation{International Centre for Theoretical Sciences, TIFR, Bangalore, India}

\author{Arnab Pal}
\email{arnabpal@imsc.res.in}
\affiliation{The Institute of Mathematical Sciences, CIT Campus, Taramani, Chennai 600113, India}
\affiliation{Homi Bhabha National Institute, Training School Complex, Anushakti Nagar, Mumbai 400094, India}


\begin{abstract}
Classical first passage under resetting is a paradigm in the search process. Despite its multitude of applications across interdisciplinary sciences, experimental realizations of such resetting processes posit practical challenges in calibrating these zero time irreversible transitions. Here, we consider a strategy in which resetting is performed using finite time return protocols in lieu of instantaneous returns. These controls could also be accompanied with random fluctuations or errors allowing target detection even during the return phase. To better understand the phenomena, we develop a unified renewal approach that can encapsulate arbitrary search processes centered around home in a fairly general topography containing targets, various resetting times and return mechanisms in arbitrary dimensions. While such finite-time protocols would apparently seem to 
prolong the overall search time in comparison to the instantaneous resetting process, we show \textit{on the contrary} that a significant speed-up can be gained by leveraging the stochasticity in home-returns. The formalism is then explored to reveal a universal criterion distilling the benefits of this strategy. We demonstrate how this general principle can be utilized to improve overall performance of a one-dimensional diffusive search process reinforced with experimentally feasible parameters. We believe that such strategies designed with inherent randomness can be made optimal with precise controllability in complex search processes. 
\end{abstract}

\pacs{Valid PACS appear here}
\maketitle

Recently, a class of non-equilibrium systems namely resetting or restart has gained a lot of attention due to their multidisciplinary applications in statistical physics \cite{evans_diffusion_2011,kusmierz2014first,gupta2014fluctuating,eule2016non,pal2015diffusion,majumdar2015dynamical,gupta2020work,biroli2023extreme,evans_stochastic_2020,gupta2022stochastic,pogorzelec2023resetting,sokolov2023linear,ghosh2023autonomous,mendez2016characterization,mori2023entropy}, chemical \& biological physics \cite{reuveni2014role,budnar2019anillin,rotbart2015michaelis,biswas2023rate,roldan2016stochastic}, quantum physics \cite{mukherjee2018quantum,yin2023restart,magoni2022emergent}, stochastic processes \cite{kumar2023universal,de2020optimization,de2022optimal,stojkoski2021geometric,wang2022entropy,bressloff2021accumulation,huang2021random}, economics \cite{jolakoski2022fate,stojkoski2022income} and operation research \cite{bonomo2021mitigating}. There has been a spate of excitement in developing theory and application of such non-equilibrium systems, and in parallel, extracting some of the rich physics that lies embedded in such systems using optical trap experiments \cite{tal2020experimental,besga2020optimal,faisant2021optimal,goerlich2023experimental} and programmable robots \cite{altshuler2023environmental,paramanick2023programming}. One of the hallmark results emanating from the seminal work of Evans and Majumdar is the ability of resetting to expedite completion of a first passage process \cite{evans_diffusion_2011}. This is noteworthy since designing optimal search processes has been a central goal of the first passage processes spanning from physics \cite{redner2001,bray2013persistence,metzler2014first}, chemistry \cite{benichou2010geometry,loverdo2008enhanced,lomholt2005optimal}, biology \cite{vergassola2007infotaxis,sheinman2012classes}, economics \cite{gabaix2016dynamics} and ecology \cite{viswanathan1999optimizing,pal_search_2020}. 

There has been a myriad of works showcasing universal dominance of resetting in diffusion alike and arbitrary stochastic processes \cite{evans_stochastic_2020,pal2022inspection,reuveni_optimal_2016,nagar2016diffusion,pal_first_2017,chechkin2018random,evans2018run,sar2023resetting,int-target-2,chen2022first,ray2021resetting,ahmad2019first,pal2019landau,belan2018restart}. Notably, a key assumption here is that resetting occurs instantaneously i.e., the system (say, a Brownian particle) can be reset (or brought back) in zero time. This is, however, a major hindrance to practical realisation or experimental verification in the field since `resetting' should be considered as a finite time process. Moreover, resetting a particle from far should require more time than to reset it from a nearby location. These presumptions seek for a more realistic viewpoint where resetting should be considered as a spatio-temporally coupled process and not just a mere teleportation. 

An important step to this direction was taken recently by considering the fact that there is a finite time return process that brings the searcher/particle back to a preferred location/home upon resetting \cite{pal_search_2020,tal2020experimental}. It was, however, assumed that the return process is deterministic with absolute precision and moreover, the searcher is blind and thus can not locate a target during the return (see also \cite{pal2019invariants,pal2019time,bodrova2020resetting,bodrova2020brownian,radice2021one,stanislavsky2022subdiffusive,bressloff2020search,bressloff2020target,maso2019transport,radice2022diffusion,mercado2020intermittent}). In realistic search processes as well as in experiments, the return motion or the driving protocols are never completely deterministic or perfect and will always be accompanied by uncontrollable random fluctuations \cite{gupta2021resetting,gupta2020stochastic,gupta2022work,mercado2020intermittent}. Importantly, precise return controls with regard to the energetics maybe costly as the thermodynamic trade-off relations have taught us \cite{barato2015thermodynamic,horowitz2020thermodynamic,pal2021thermodynamic,pal2023thermodynamic}. It is therefore important to understand how the errors or fluctuations in return process can be incorporated and secondly, their ramifications to the overall search.


To this end, we develop a unified framework of search process with stochastic home return that can access the targets during the return phase. We obtain the mean search time for an arbitrary search and return process in $d$-dimension.
Quite remarkably, we find that such innate randomness or errors in return times, albeit adding overhead time penalties in general to the overall mean search time, can \textit{often} expedite the classical first passage process under instantaneous resetting that takes no penalty. To understand this physically, we unravel a universal trade-off relation between two timescales emanating from pre- and post-resetting phases respectively.
Key features relating to this global advantage are then highlighted for a paradigmatic diffusive search process that can be realized in optical trap experiments.

\begin{figure}[t!]
    \centering
    \includegraphics[width=8cm]{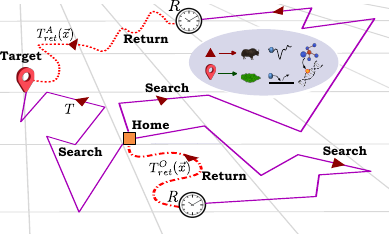}
    \caption{Schematic diagram of an agent (could be a foraging animal, colloid or a bio-molecule) searching for targets (could be resources, sticky surface or a defected site along DNA). The agent can find the target during the \textit{search phase} (violet solid trajectory). Otherwise, after a random resetting time $R$ the agent, currently located at $\Vec{x}$, switches to a return phase (facilitated e.g., by a chemical or potential gradient or simply, by choice) when it either returns to its home or relocates to a new location in finite time $T_{ret}^O(\Vec{x})$ (red dot-dashed trajectory). The agent also has a finite probability to find the target during its return phase (red dotted trajectory) in time $T_{ret}^A(\Vec{x})$. Target detection marks the completion of the global search regardless of its phase. We ask: Whether search with a finite-time home return can perform better than the instantaneous return process? }
    \label{fig1}
\end{figure}

\textbf{\emph{General framework.---}} Consider a prototypical random search process where a searcher starts at the origin $O$, its home, of a $d$-dimensional arena at time zero (Fig. \ref{fig1}). The arena may contain one or multiple targets. The searcher can locate one of these targets following a random $T$ which we denote as the first passage time. However, during the search phase, if the target is not found up to a random time $R$ (which we denote as the resetting time), the searcher decides to return to its home. Crucially, the return process can be stochastic and it is assumed moreover that the searcher has the ability to detect the target(s) even during the return phase. If the searcher returns to $O$, failing to detect the target, the search process restarts afresh to the next attempt. Overall, the process is said to be completed once the target is detected irrespective of the searcher's phase (see Fig. \ref{fig1}).

The time it takes for the searcher to either find the target or the origin (home) during return will typically depend on the searcher’s position $\Vec{x}$ at the time of resetting $R$. Furthermore, various demographical or physical constraints may compel the searcher to follow different routes hence there can be more nontrivial dependence between the position of the searcher and the stochastic return time either to the origin (denoted by $T_{ret}^O(\Vec{x})$) or to any of the targets $A$ (denoted by $T_{ret}^A(\Vec{x})$). In the former case, the searcher resumes its search phase upon returning to the origin while in the latter case the process ends. Denoting the overall completion time by a random variable $T_R$ and considering the above-mentioned possibilities, we can write a renewal equation as follows
\begin{align}
\begin{array}{l}
T_{R}=\left\{ \begin{array}{lll}
T \hspace{3cm} &\text{if }T<R\\
R+T_{ret}^A(\Vec{x})\hspace{1.5cm} & \text{if }R \leq T~ \& ~T_{ret}^A(\Vec{x})<T_{ret}^O(\Vec{x})\\
 R+T^O_{ret}(\Vec{x})+T_R'\hspace{0.6cm} &\text{if }R \leq T ~
\&~ T_{ret}^O(\Vec{x})\leq T_{ret}^A(\Vec{x}) \end{array}\right.\text{ }\end{array}
\label{renewal}
\end{align}
where we have assumed that return will be prioritized if $R=T$. Here, $T_R'$ is a statistically independent and identically distributed copy of $T_R$. \eref{renewal} can be written in a more concise form as
\begin{align}
    T_R&=min(T,R)+I(R\le T) min(T_{ret}^A(\Vec{x}), T_{ret}^O(\Vec{x})) \nonumber \\
    &+I(R\le T)I(T_{ret}^O(\Vec{x})\le T_{ret}^A(\Vec{x}))T_R',
        \label{S2}
\end{align}
where $min(u,v)$ is the minimum of two random variables $u$ \& $v$ and $I(u  \leq v)$ is an indicator function that takes value unity when $u \leq v$, and zero otherwise. Thus, $ \langle I(u  \leq v) \rangle=Pr(u \leq v)$ i.e, the probability that $u \leq v$.  Crucially, the indicator function $I(T_{ret}^O(\Vec{x})\le T_{ret}^A(\Vec{x}))$ also depends on the coordinate $\Vec{x}$ of the searcher at the time of resetting. This implies that the expectations on \eref{S2} need to be performed over the underlying stochastic process, the resetting time density $f_R(t)$ and the return protocols. For instance, the expectation $\mathcal{E} \equiv \langle I(R\le T)I(T_{ret}^O(\Vec{x})\le T_{ret}^A(\Vec{x})) T_R'\rangle$ can be computed as follows (see Sec S1A in \cite{SI})

{\footnotesize
\begin{align}
    \mathcal{E}&= \int_0^\infty dt f_R(t)  \frac{Pr(T\ge t) \int_{\mathcal{D}}d\Vec{x}G(\Vec{x},t)\langle I(T_{ret}^O(\Vec{x})\le T_{ret}^A(\Vec{x})) \rangle}{Q(t)} 
 \langle T_R' \rangle \nonumber \\
 &= \langle T_R \rangle \int_0^\infty dt f_R(t) \int_{\mathcal{D}}d\Vec{x}G(\Vec{x},t) Pr(T_{ret}^O(\Vec{x})\le T_{ret}^A(\Vec{x})),
 \label{expt-1}
\end{align}}where $\mathcal{D}$ is the domain of search in arbitrary dimension that can contain one or multiple targets and $G(\Vec{x},t)$ is the underlying time-dependent propagator in the presence of targets so that $Q(t)=Pr(T\geq t) =\int_{\mathcal{D}}d\Vec{x}~G(\Vec{x},t)$ becomes the survival probability \cite{redner2001}. In the first line of Eq. (\ref{expt-1}), we have taken averages over $f_R(t)$ and the underlying search propagator. Computing similar expectations from Eq. (\ref{S2}), we obtain the mean first passage time (MFPT) to be \cite{SI}
\begin{align}
\langle T_R \rangle =\frac{\langle min(T,R) \rangle +\langle min\left(T_{ret}^A,T_{ret}^O\right) \rangle}{1-\int_{\mathcal{D}}d\Vec{x}~\widetilde{G}_R(\Vec{x}) Pr(T_{ret}^O(\Vec{x})\le T_{ret}^A(\Vec{x}))} ,
     \label{mfpt}
\end{align}
where $\widetilde{G}_R({\Vec{x}}) \equiv \int_0^\infty G(\Vec{x},t)f_R(t)dt $ is the time-integrated propagator and $\langle min\left(T_{ret}^A,T_{ret}^O\right) \rangle=\int_{\mathcal{D}}d\Vec{x}\widetilde{G}_R(\Vec{x})\langle min\left(T_{ret}^A(\Vec{x}),T_{ret}^O(\Vec{x})\right)\rangle$. On the other hand,  $Pr(T_{ret}^O(\Vec{x})\le T_{ret}^A(\Vec{x}))$ is the splitting probability that the searcher, starting from $\Vec{x}$, reaches the origin before hitting any of the targets during the return phase.

\eref{mfpt} is central to this study and is remarkably general since it holds for a) any kind of underlying first passage process  (beyond diffusion) conducted in any dimension in the presence of arbitrary target(s) regardless of their variation in size, shape or nature (purely absorbing or partial), b) arbitrary resetting time distributions, and c) generic returning motion such as instantaneous, deterministic or stochastic and their various modes of return.

\textbf{\emph{A universal criterion for the trade-off between instantaneous and stochastic return---} }
 Various optimization questions could be asked given the generality of the formalism. Specifically, we ask whether stochastic returns can over-perform the search compared to the instantaneous returns irrespective of the underlying process. Our analysis shows that stochastic return will be beneficial only if 
$    \langle T_R \rangle < \langle T_R^{inst} \rangle$, where $\langle T_R^{inst} \rangle=\frac{\langle min(T,R) \rangle}{Pr(T<R)}$ is the MFPT for the instantaneous return \cite{pal_first_2017} and can be obtained by noting that the searcher always returns to the origin in zero time so that $Pr(T_{ret}^O\le T_{ret}^A)=1$. Evidently, the condition above yields
\begin{align}
\mathcal{T} \equiv 
\frac{
\langle min\left(T_{ret}^A,T_{ret}^O\right) \rangle}{Pr(T_{ret}^A<T_{ret}^O)} < \langle T_R^{inst} \rangle ,
  \label{srcond}
\end{align}
where $Pr(T_{ret}^A<T_{ret}^O)=\int_{\mathcal{D}}d\Vec{x}~\widetilde{G}_R(\Vec{x}) Pr(T_{ret}^A(\Vec{x})<T_{ret}^O(\Vec{x})) $. 
Clearly, to understand this trade-off, one needs to consider many realizations of such a process and compare between the time $\mathcal{T}$ it takes, on an average, for the process to either return to the origin or to complete the search, starting from the location at the time of resetting given by $\langle min\left(T_{ret}^A,T_{ret}^O\right) \rangle$ (rescaled with the splitting probability $Pr(T_{ret}^A<T_{ret}^O)$) and the time $ \langle T_R^{inst} \rangle$ that it would have taken for the instantaneous return. This is rather intriguing since the former accumulates time only during the post resetting return phase while latter does so only during the pre-resetting phase. Thus, the relation (\ref{srcond}) puts a strong constraint on the average return time $\mathcal{T}$ regardless of the final destination. Notably, this relation is quite universal for it does not depend on the particular choice of the underlying first passage process, resetting time density 
or the return protocol. Furthermore, the relation can be used as a guiding principle to design search-efficient protocols as will be shown below.

\textbf{\emph{Diffusive search with resetting and stochastic return via controlled potential trap.---}} To demonstrate the power of our approach, we examine the paradigm of a 1d diffusive search process (characterised by the diffusion constant $D$) in which
a particle starts at the origin $O$ and continues to diffuse until it hits a
stationary target at a location $L$. In addition, assume that the process is reset at a constant rate $r$ (i.e., resetting time density $f_R(t)=re^{-rt}$) upon which a potential $U(x)$ centered at the origin is turned on. The particle diffuses through the potential and it is switched off when the particle makes a first return to the origin. Subsequently, the particle resumes its diffusive search phase (see \cite{gupta2021resetting,gupta2020stochastic} where non-equilibrium steady state and the relaxation properties were studied under this protocol). However, since the return is not purely deterministic, there is always a chance to find the target during the return phase (Fig. \ref{fig2} inset). In below, we demonstrate how this could expedite the overall search time.

We start by recalling the diffusive propagator of this
process given by $ G(x,t)=\frac{1}{\sqrt{4\pi Dt}}\left(e^{-\frac{x^2}{4Dt}} -e^{-\frac{(2L-x)^2}{4Dt}}\right)$ \cite{redner2001}. For diffusive search process, the underlying first passage times $(T)$ are sampled from the L\'evy Smirnov distribution so that $\langle min(T,R) \rangle=\frac{1}{r} \left(1- e^{-\sqrt{rL^2/D}} \right)$ \cite{pal_first_2017}. During the return phase, the particle has the possibility to hit either $O$ or $L$ starting from position $x$ which is the 
coordinate of the particle exactly at the time of resetting. 
Considering that the return phase is facilitated by a linear potential $U(x)=\lambda |x|$, we can compute the average time 
for the particle
to reach either of the boundaries namely $\langle t_{2}(x) \rangle = \frac{L(1-e^{\lambda x/D}) +x(e^{\lambda L/D }-1)}{\lambda(e^{\lambda L/D}-1)} $ for $x>0$ and $\langle t_{1}(x) \rangle =|x|/\lambda$ for $x<0$ \cite{redner2001}. Using this one can write the following expectation \cite{SI}
\begin{eqnarray}
\langle min\left(T_{ret}^L(x),T_{ret}^O(x)\right)\rangle=   \theta(-x)\langle t_1(x) \rangle + \theta(x)\langle t_2(x) \rangle,
\end{eqnarray}
where $\theta(x)$ is the step-function. In addition, the splitting probability is given by given by $Pr(T_{ret}^O(x)\le T_{ret}^L(x))=\left[ \theta(x)\frac{e^{\lambda x/D}-e^{\lambda L/D}}{1-e^{\lambda L/D}}+\theta(-x) \right]$ \cite{redner2001}. Substituting these expressions into \eref{mfpt}, we arrive at the following expression for the MFPT  $\langle T_R \rangle=D/L^2 \langle \tau(\overline{r},\overline{\lambda}) \rangle$, where
\begin{small}
\begin{eqnarray}
     &&\langle \tau (\overline{r}, \overline{\lambda}) \rangle = \frac{1}{\overline{r} \overline{\lambda}^2}\bigg(2 \overline{\lambda}^2 +2 e^{\overline{\lambda}} \overline{r} -2 e^{\overline{\lambda}} \overline{\lambda}^2-\overline{r} \overline{\lambda} -2 \overline{r} \nonumber \\
     &&\hspace{0.4cm} +\frac{2 \left(e^{\overline{\lambda}}-1\right) \overline{\lambda} \left(\left(2 e^{\overline{\lambda}}-1\right) e^{\sqrt{\overline{r}}}-1\right) \left(\overline{\lambda}^2-\overline{r}\right)}{2 e^{\overline{\lambda}+\sqrt{\overline{r}}}\overline{\lambda} -e^{2 \sqrt{\overline{r}}} \left(\overline{\lambda}+\sqrt{\overline{r}}\right)-\overline{\lambda}+\sqrt{\overline{r}}}\Bigg),
     \label{mfpt1}    
 \end{eqnarray}
 \end{small}
and $\overline{r}=\frac{rL^2}{D}, \overline{\lambda}=\frac{\lambda L}{D}$ are the scaled (dimensionless) resetting rate and potential strength respectively. \fref{fig2} shows corroborated plots (theory \& simulations) of $\langle \tau (\overline{r}, \overline{\lambda}) \rangle$ as a function of $\overline{\lambda}$. \eref{mfpt1} reproduces $ \langle \tau_{inst}(\overline{r}) \rangle=\langle \tau (\overline{r},\overline{\lambda} \to \infty) \rangle =  \frac{e^{\sqrt{\overline{r}}}-1}{\overline{r}}$ which is the MFPT for instantaneous resetting \cite{evans_diffusion_2011} -- shown in Fig. \ref{fig2} by the horizontal dashed lines for different resetting rates.

Intriguingly, Fig. \ref{fig2} shows a key feature that
the MFPT for the stochastic return can be reduced further than the instantaneous resetting (see e.g., the $\overline{r}=10$ curve). In fact, \eref{mfpt1} reveals the existence of a critical resetting rate $\overline{r}^*$, which is a function of $\overline{\lambda}$, beyond which stochastic return always supersedes the instantaneous return in optimizing the search time (see later for the exact evaluation of $\overline{r}^*$). However, for $\overline{r}<\overline{r}^*$ (see e.g., the $\overline{r}=5$ curve), the MFPT always stays above the instantaneous (dashed) line indicating that no advantage can be gained from the stochastic return for any resetting rate or $\overline{\lambda}$.

\begin{figure}
    \centering
    \includegraphics[width=8cm]{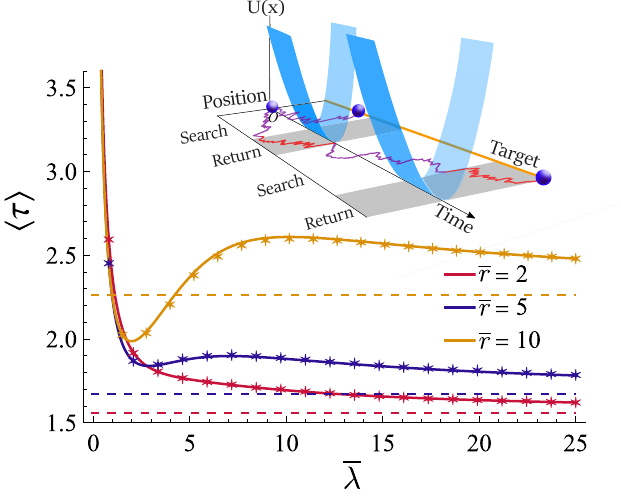}
    \caption{Variation of the MFPT $\langle \tau(\overline{r},\overline{\lambda}) \rangle $ from (\ref{mfpt1}), shown by the solid curves, as a function of $\overline{\lambda}$ for different resetting rate $\overline{r}$. The horizontal dashed lines represent $\langle \tau_{inst}(\overline{r}) \rangle$ for respective $\overline{r}$. Evidently, stochastic return can both reduce ($\overline{r}=10$) or prolong ($\overline{r}=5$) the completion time compared to the instantaneous returns. This behavioural transition occurs at $\overline{r}^*$ which is strictly a function of $\overline{\lambda}$. Numerical data (depicted by the stars) shows an excellent agreement with the theory. Inset schematic: Possible trajectories of a diffusive searcher under resetting and stochastic return facilitated by a potential trap. Violet (red) trajectory corresponds to search \& trap-off (return \& trap-on) phase.}
    \label{fig2}
\end{figure}


\textbf{\textit{A universal phase diagram for diffusive search.}--}
The inequality in Eq. (\ref{srcond}) allows us to construct a universal phase diagram, spanned by the system parameters, that govern the dominance of stochastic return over the classical instantaneous return. Such a phase diagram is graphically illustrated in \fref{fig5} for the 1d diffusion. The red dashed line, obtained by setting $\mathcal{T}=\langle T_R^{inst} \rangle$, separates the two phases namely stochastic return -`beneficial' and -`detrimental' than the instantaneous return. The separatrix is the locus of the set of critical resetting rate $\overline{r}^*$ for each $\overline{\lambda}$ obtained from the above equality (see \cite{SI} for further elaborations).

To gain further insights into the trade-off between two phases, we study $\mathcal{T}$ and discuss some of the limiting cases. For \textit{very low resetting rate} ($\overline{r}\to0$) and finite $\overline{\lambda}$, one has  $\mathcal{T} \sim \frac{1}{\overline{r}} \left( \frac{2 \overline{\lambda} \left(1-e^{\overline{\lambda}}\right)  }{\overline{\lambda} (\overline{\lambda}+2)-2 e^{\overline{\lambda}}+2  } \right) $ whereas $\langle\tau_{inst}(\overline{r}) \rangle \propto 1/\sqrt{\overline{r}}$. Clearly, the LHS diverges faster than RHS nullifying the criterion (\ref{srcond}) for any $\overline{\lambda}$. 
Indeed, the trajectories that go away from the target accumulate more time during the stochastic return than the instantaneous return making the former strategy detrimental.
 In contrast for $\overline{r}\gg\mathcal{O}(1)$ - \textit{frequent resets} - and finite $\overline{\lambda}$, the RHS diverges as $\propto e^{\sqrt{\overline{r}}}/\overline{r}$ while the LHS is bounded as $\mathcal{T}=\frac{2 e^{\overline{\lambda}}-\overline{\lambda}-2}{\overline{\lambda}^2}$ which is identical to $\langle \tau(\overline{r}\to \infty,\overline{\lambda}) \rangle$. Here, the pre-resetting phase is very short and the particle is effectively in the return phase. Thus, it has a finite probability to find the target during return phase while the instantaneous return almost always keeps the particle close to the origin eliminating the target-detection. Clearly, this is a favourable situation for the stochastic returns. 
 

For the intermediate case of \textit{finite $\overline{r}$}, there are two limiting cases of $\overline{\lambda}$. For large $\overline{\lambda}$, one finds $\mathcal{T} \sim \overline{\lambda}\left(\frac{2  \sinh \left(\sqrt{\overline{r}}\right)}{\overline{r}^{3/2}}-\frac{1}{\overline{r}}\right)$ to be divergent while the RHS is finite. Note that in this limit the return probability to the origin (having return time $\approx \sqrt{D/r}/\lambda$) is almost close to unity and the return phase only adds time penalties. Naturally, instantaneous returns are more efficient. For small $\overline{\lambda}$, however, the average return time is large (specifically from the trajectories in the $x<0$ region) as can be seen from $\mathcal{T}\sim\frac{1}{\overline{\lambda}}\frac{1}{1-\sqrt{\overline{r}} \text{csch}\left(\sqrt{\overline{r}}\right)}\propto\frac{1}{\overline{\lambda}}$ which diverges invalidating the condition (\ref{srcond}). This ensures that instantaneous returns are more beneficial. In the bulk, the competing effect is non-trivial due to the intricate interplay between $\overline{r}$ and $\overline{\lambda}$ (see Sec S10 and Table 1 in \cite{SI}). 


 \begin{figure}[t]
    \centering
    \includegraphics[width=8.7cm]{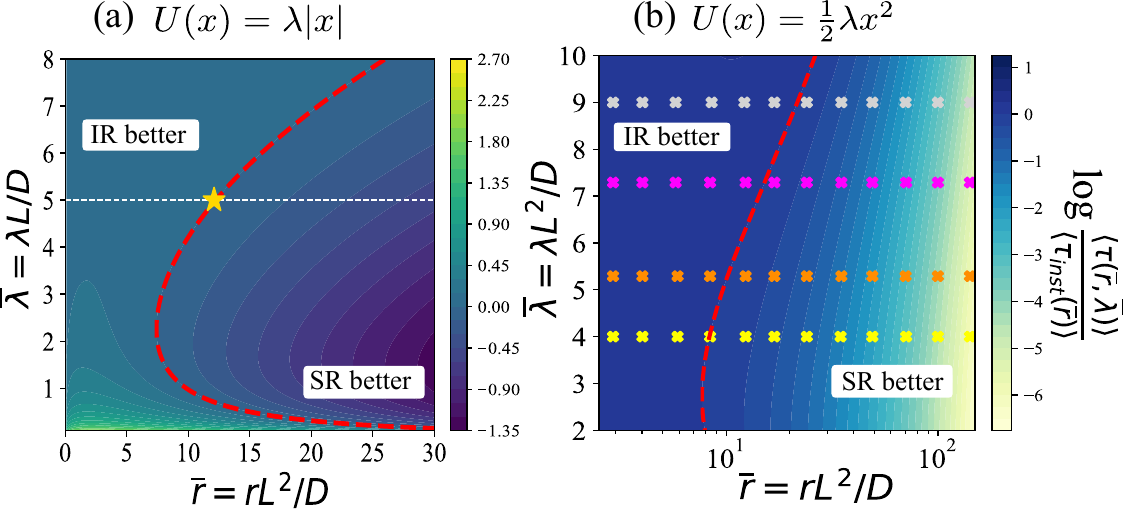}
    \caption{Phase diagram for diffusive search with stochastic returns driven by linear- (panel a) and harmonic- (panel b) potentials: The region right to the phase boundary (red dashed curve) is where condition (\ref{srcond}) is satisfied and stochastic return (SR) supersedes instantaneous return (IR). The red curve is obtained by setting  (\ref{srcond}) to an equality. The side colorbar is an estimation of search efficiency, defined by the ratio $\frac{\langle \tau(\overline{r},\overline{\lambda}) \rangle}{\langle \tau_{inst}(\overline{r}) \rangle}$; values in log-scale - positive (negative) for slow-down (speed-up) by the stochastic return. In panel (a), the yellow star indicates the critical $\overline{r}^*$ which is obtained from the intersection of the separatrix and the horizontal white line spanned by each $\overline{\lambda}$ (here, $\overline{\lambda}=5,~\overline{r}^*\approx 12$). In panel (b), we have superimposed the data points (marked as crosses) from the experimental parameter sets $\left( \overline{\lambda}, \overline{r} \right)$ from \cite{besga2020optimal} (details in \cite{SI}).}
    \label{fig5}
\end{figure}


\textbf{\textit{A pragmatic application of the phase diagram.---}} Let us now consider a 1d diffusive search process where returns are facilitated by a harmonic trap $U(x)=\frac{1}{2}\lambda x^2$ (replacing the linear trap) that can be easily fabricated in experiments \cite{besga2020optimal,faisant2021optimal}. In this case, the phase diagram spanning in the parameter space $ \left( \overline{\lambda}=\frac{\lambda L^2}{D}, \overline{r}=\frac{r L^2}{D} \right)$ demonstrates that stochastic returns are indeed more efficient than instantaneous returns for a range of parameters as can be seen from Fig. \ref{fig5}b. Importantly here, variation of the parameters is not arbitrary, but we have used a list of the parameters $\left( \overline{\lambda}, \overline{r} \right)$ that were used in the experiments \cite{besga2020optimal,faisant2021optimal} (we refer to S11 in \cite{SI} for the detailed analysis of the MFPT, the connection between theory \&  experiments and the phase diagram). Thus, in an attempt to verify our results (with proposed stochastic return protocols) experimentally in similar set-ups as in \cite{besga2020optimal,faisant2021optimal}, Fig. \ref{fig5}b would be the guiding phase diagram to selectively choose parameters for showcasing various trade-offs. Finally, we emphasize that the phase diagram also quantifies the speed-up in search efficiency that can be gained by implementing stochastic return protocols for a range of experimentally accessible parameters.


\textbf{\emph{Conclusions.---}}
In this letter, we have developed a unified first passage time framework of a stochastic search process under finite time resetting or returns. Notable distinction between this and the existing body of works \cite{evans_diffusion_2011,bodrova2020resetting,pal_search_2020,gupta2021resetting} lies on the fact that the home-returns can be accompanied with stochasticity and thus the searchers can be fortuitous to find targets during the return. Naively, one expects that finite time returns can only incur delay to the overall completion time. However, we argue that the element of randomness in return process
can expedite the overall completion even in comparison to the classical instantaneous resetting process which takes no time to return. This is the most intriguing observation of this work to which we attribute various physical scenarios. Elucidating further the scope of such observation to arbitrary stochastic process with generic returns, we derive a universal and physically amenable criterion that unveils the superiority of the finite time stochastic returns above the zero-time returns. We 
emphasize that the universal framework of this problem is also a powerful tool to predict the fluctuations and possibly the full distribution. 

We believe that resetting with stochastic home returns can turn out to be a universal optimization strategy owing to its dominance over classical first passage resetting with applications to biochemical search \cite{benichou2011intermittent,iyer2016first} and molecular transport \cite{jain2023fick,metzler2014first}. Whether stochastic return can be beneficial than instantaneous return for an optimally restarted process is a challenging question that will be discussed elsewhere.
Importantly, our approach is potentially useful for conducting experiments from the practical implementation of resetting especially since one need not drag or track the agents (e.g., colloids or programmable robots) all the way during the return process \cite{besga2020optimal}. Finally, it is intuitive that stochastic return could be energetically optimal while considering the search completion than the deterministic return protocols. These frontiers remain to be explored further in future.

\emph{\textbf{Acknowledgements.}---}  The numerical calculations reported in
this work were carried out on the Nandadevi cluster, which is maintained and supported by the Institute of Mathematical Science’s High-Performance Computing Center. AK acknowledges the support of the core research grant CRG/2021/002455
and the MATRICS grant MTR/2021/000350 from the SERB, DST, Government of India. AK also acknowledges support of the Department of Atomic Energy, Government of India, under Project No. RTI4001. AP gratefully acknowledges research support from the Department of Science and Technology, India, SERB Start-up Research Grant Number SRG/2022/000080 and Department of Atomic Energy, Government of India.

\bibliography{fpusr}

\begin{titlepage}
\title{Supplemental Material for \\``Stochasticity in returns can expedite classical first passage under resetting''}
\maketitle
\end{titlepage}

\onecolumngrid
\setcounter{page}{1}
\renewcommand{\thepage}{S\arabic{page}}
\setcounter{equation}{0}
\renewcommand{\theequation}{S\arabic{equation}}
\setcounter{figure}{0}
\renewcommand{\thefigure}{S\arabic{figure}}
\setcounter{section}{0}
\renewcommand{\thesection}{S\arabic{section}}
\setcounter{table}{0}
\renewcommand{\thetable}{S\arabic{table}}

This Supplemental Material provides detailed mathematical derivations and additional discussions which support the results described in the Letter. Moreover, it also provides details of the examples used in the main text to demonstrate the validity and applicability of the formalism.

\tableofcontents

\section{Derivation of the MFPT (Eq. 4) in the main text}\label{sc1}
In this section, we will provide a detailed derivation of Eq. (4) -- the mean first passage time (MFPT). We start from the stochastic renewal Eq. (1) 
\begin{align}
\begin{array}{l}
T_{R}=\left\{ \begin{array}{lll}
T \hspace{3cm} &\text{if }T<R\\
R+T_{ret}^A(\Vec{x})\hspace{1.5cm} & \text{if }R \leq T~ \& ~T_{ret}^A(\Vec{x})<T_{ret}^O(\Vec{x})\\
 R+T^O_{ret}(\Vec{x})+T_R'\hspace{0.6cm} &\text{if }R \leq T ~
\&~ T_{ret}^O(\Vec{x})\leq T_{ret}^A(\Vec{x}) \end{array},\right.\text{ }\end{array}
\label{renewal1}
\end{align}
which can be written in a concise form as
\begin{align}
    T_R&=min(T,R)+I(R\le T) min(T_{ret}^A(\Vec{x}), T_{ret}^O(\Vec{x}))
    +I(R\le T)I(T_{ret}^O(\Vec{x})\le T_{ret}^A(\Vec{x}))T_R',
        \label{renewal-2}
\end{align}
where $min(u,v)$ is the minimum of two random variables $u$ \& $v$ and $I(u  \leq v)$ is an indicator function that takes value unity when $u \leq v$, and zero otherwise. Thus, $ \langle I(u  \leq v) \rangle=Pr(u \leq v)$ i.e, the probability that $u \leq v$. Using the definition of the random variable $min(u,v)$, we can write 
\begin{align}
    min(T_{ret}^A(\Vec{x}), T_{ret}^O(\Vec{x}))=I(T_{ret}^A(\Vec{x})< T_{ret}^O(\Vec{x}))T_{ret}^A(\Vec{x})+I(T_{ret}^O(\Vec{x})\le T_{ret}^A(\Vec{x}))T_{ret}^O(\Vec{x}).
    \label{min}
\end{align} 
We now take expectations on the both sides of \eref{renewal-2} to have
\begin{align}
    \langle T_R \rangle&= \langle min(T,R) \rangle + \langle I(R\le T) min(T_{ret}^A(\Vec{x}), T_{ret}^O(\Vec{x})) \rangle + \langle I(R\le T)I(T_{ret}^O(\Vec{x})\le T_{ret}^A(\Vec{x}))T_R' \rangle.
    \label{s1}
\end{align}
Noting that $T_R'$ is an independent and identically distributed copy of $T_R$, we can simplify the last expression from \eref{renewal-2} as $ \langle I(R\le T)I(T_{ret}^O(\Vec{x})\le T_{ret}^A(\Vec{x})) \rangle \langle T_R' \rangle=\langle I(R\le T)I(T_{ret}^O(\Vec{x})\le T_{ret}^A(\Vec{x})) \rangle \langle T_R \rangle $.  Rearranging the terms in \eref{renewal-2} then gives us
\begin{align}
    \langle T_R \rangle &= \underbrace{\frac{\langle min(T,R) \rangle}{1-\langle I(R\le T)I(T_{ret}^O(\Vec{x})\le T_{ret}^A(\Vec{x})) \rangle}}_{\text{search/exploration phase}} +\underbrace{\frac{\langle I(R\le T)I(T_{ret}^O(\Vec{x})\le T_{ret}^A(\Vec{x}))T_{ret}^O(\Vec{x}) \rangle}{1-\langle I(R\le T)I(T_{ret}^O(\Vec{x})\le T_{ret}^A(\Vec{x})) \rangle}}_{\text{return phase: return to origin}} \nonumber \\
    &\hspace{7cm}+ \underbrace{\frac{\langle I(R\le T)I(T_{ret}^A(\Vec{x})< T_{ret}^O(\Vec{x}))T_{ret}^A(\Vec{x}) \rangle}{1-\langle I(R\le T)I(T_{ret}^O(\Vec{x})\le T_{ret}^A(\Vec{x})) \rangle}}_{\text{return phase: finding the target}}
    \label{s22}.
\end{align}
The first term in \eref{s22} accounts for the time taken due to the exploration of the searcher.  The second term contributes to the time that the searcher takes to return to the starting position. Finally, the third term accounts for the time spent while finding the target during the return phase. It should be noted that the expectations in \eref{s22} are taken over three different random components: the underlying stochastic process, the resetting time, and the stochastic return phase trajectories. In what follows, we show how to evaluate these expectations.

\subsection{The denominator}
First we focus on the term in the denominator that is common to all the components. Conditioned on the resetting time $R$ over the density $f_R(t)$, we can write
\begin{align}
   \langle I(R\le T)I(T_{ret}^O(\Vec{x})\le T_{ret}^A(\Vec{x})) \rangle  &=\int_0^\infty dt f_R(t) Pr(T\ge t) \langle I(T_{ret}^O(\Vec{x}(t))\le T_{ret}^A(\Vec{x}(t))) \rangle_{\Vec{x},return},
   \label{s3}
\end{align}
where further averaging needs to be done over the underlying process $\Vec{x}$ and then the return process (denoted by the subscripts). The former is done by noting that at the time of resetting, the searcher could be anywhere in the domain $\mathcal{D}$ and thus its position is sampled over $G(\Vec{x},t)$ -- the propagator in the presence of targets. Evidently, the search upto resetting time is possible only if the searcher was not absorbed by then -- this occurs essentially with the survival probability $Q(t)$
given by 
\begin{align}
    Q(t)=Pr(T\ge t)=\int_{\mathcal{D}}d\Vec{x}~G(\Vec{x},t).
    \label{surv}
\end{align}
Thus the averaging should be over the normalized PDF $G(\Vec{x},t)/Q(t)$ so that the expectation inside the integral in \eref{s3} reads
\begin{align}
     &\int_0^\infty dt f_R(t) Pr(T\ge t) \langle I(T_{ret}^O(\Vec{x}(t))\le T_{ret}^A(\Vec{x}(t))) \rangle_{\Vec{x},return} \nonumber\\
     =&\int_0^\infty dt f_R(t) \cancel{Pr(T\ge t) }\frac{\int_{\mathcal{D}}d\Vec{x}G(\Vec{x},t)\langle I(T_{ret}^O(\Vec{x})\le T_{ret}^A(\Vec{x})) \rangle_{return}}{\cancel{\int_{\mathcal{D}}d\Vec{x}G(\Vec{x},t)}} \nonumber \\
     =& \int_0^\infty dt f_R(t) \int_{\mathcal{D}}d\Vec{x}G(\Vec{x},t)\langle I(T_{ret}^O(\Vec{x})\le T_{ret}^A(\Vec{x})) \rangle.
     \label{s4}
\end{align}
Here we have omitted the subscript denoting average over return motion as this is obvious. We shall not use the subscripts in the following derivation as these same notations follow there. We are now left with the last expectation on the return process which is straightforward such that
 \begin{align}
 \langle I(R\le T)I(T_{ret}^O(\Vec{x})\le T_{ret}^A(\Vec{x})) \rangle & = \int_0^\infty dt f_R(t) \int_{\mathcal{D}}d\Vec{x}~G(\Vec{x},t)\langle I(T_{ret}^O(\Vec{x})\le T_{ret}^A(\Vec{x})) \rangle \nonumber \\
& = \int_0^\infty dt f_R(t) \int_{\mathcal{D}}d\Vec{x}~G(\Vec{x},t)Pr(T_{ret}^O(\Vec{x})\le T_{ret}^A(\Vec{x})) .
\label{s5}
\end{align}
The above-mentioned steps were used to derive the expectation $\mathcal{E} \equiv \langle I(R\le T)I(T_{ret}^O(\Vec{x})\le T_{ret}^A(\Vec{x})) T_R'\rangle$  [Eq. (3)]
in the main text. Furthermore, introducing the time-integrated propagator 
\begin{align}
  \widetilde{G}_R({\Vec{x}})= \int_0^\infty G(\Vec{x},t)f_R(t)dt   ,
 \end{align}
 we can rewrite \eref{s5} as
\begin{align}
     \langle I(R\le T)I(T_{ret}^O(\Vec{x})\le T_{ret}^A(\Vec{x})) \rangle  = \int_{\mathcal{D}}d\Vec{x}~\widetilde{G}_R(\Vec{x}) Pr(T_{ret}^O(\Vec{x})\le T_{ret}^A(\Vec{x})).
\end{align}
In \eref{s5}, the term $Pr(T_{ret}^O(\Vec{x})\le T_{ret}^A(\Vec{x}))$ can be interpreted as the splitting probability of the searcher to $O$. Simply put, this is the probability that the particle has first reached $O$ without touching any of the other boundaries.

\subsection{The second term} 
For the numerator of the second term in \eref{s22} we proceed with similar arguments as in the previous subsection. This boils down to
\begin{align}
    &\langle I(R\le T)I(T_{ret}^O(\Vec{x})\le T_{ret}^A(\Vec{x}))T_{ret}^O(\Vec{x}) \rangle \\
    =& \int_0^{\infty}dt f_R(t) Pr(T\ge t) \langle I(T_{ret}^O(\Vec{x}(t))\le T_{ret}^A(\Vec{x}))T_{ret}^O(\Vec{x}(t)) \rangle\nonumber\\
    =& \int_0^\infty dt f_R(t) \cancel{Pr(T\ge t) }\frac{\int_{\mathcal{D}}d\Vec{x}G(\Vec{x},t)\langle I(T_{ret}^O(\Vec{x})\le T_{ret}^A(\Vec{x})) T_{ret}^O(\Vec{x})\rangle}{\cancel{\int_{\mathcal{D}}d\Vec{x}G(\Vec{x},t)}}\nonumber\\
    =&\int_{\mathcal{D}}d\Vec{x}\widetilde{G}_R(\Vec{x})\langle I(T_{ret}^O(\Vec{x})\le T_{ret}^A(\Vec{x}))T_{ret}^O(\Vec{x})\rangle \nonumber \\
    =&\int_{\mathcal{D}}d\Vec{x}\widetilde{G}_R(\Vec{x}) Pr(T_{ret}^O(\Vec{x})\le T_{ret}^A(\Vec{x}))\langle T_{ret}^O(\Vec{x})|T_{ret}^O(\Vec{x})\le T_{ret}^A(\Vec{x})\rangle,
    \label{s6}
\end{align}
where $\langle T_{ret}^O(\Vec{x})|T_{ret}^O(\Vec{x})\le T_{ret}^A(\Vec{x})\rangle$ is the conditional mean first passage time of the searcher to reach the origin $O$ before finding any of the targets during the return phase. This will mark the completion of the return phase following which the search/exploration phase will resume.

\subsection{The third term}
Following the same procedure as before, we have 
\begin{align}
& \langle I(R\le T)I(T_{ret}^A(\Vec{x})< T_{ret}^O(\Vec{x}))T_{ret}^A(\Vec{x}) \rangle \nonumber\\
=&\int_0^{\infty} dt f_R(t) Pr(T\ge t) \langle I(T^A_{ret}(\Vec{x}(t))<T^O_{ret}(\Vec{x}(t))) T^A_{ret}(\Vec{x}(t)) \rangle \nonumber\\
 =& \int_0^\infty dt f_R(t) \cancel{Pr(T\ge t) }\frac{\int_{\mathcal{D}}d\Vec{x}G(\Vec{x},t)\langle I(T_{ret}^A(\Vec{x})< T_{ret}^O(\Vec{x})) T_{ret}^A(\Vec{x})\rangle}{\cancel{\int_{\mathcal{D}}d\Vec{x}G(\Vec{x},t)}} \nonumber\\
 =&\int_{\mathcal{D}}d\Vec{x}\widetilde{G}_R(\Vec{x})\langle I(T_{ret}^A(\Vec{x})< T_{ret}^O(\Vec{x}))T_{ret}^A(\Vec{x})\rangle \nonumber \\
 =&\int_{\mathcal{D}}d\Vec{x}\widetilde{G}_R(\Vec{x}) Pr(T_{ret}^A(\Vec{x})< T_{ret}^O(\Vec{x}))\langle T_{ret}^A(\Vec{x})|T_{ret}^A(\Vec{x})< T_{ret}^O(\Vec{x})\rangle,
 \label{s7}
 \end{align}
where $\langle T_{ret}^A(\Vec{x})|T_{ret}^A(\Vec{x})< T_{ret}^O(\Vec{x})\rangle$ is the conditional mean first passage time of the searcher to find any of the targets before reaching the origin during the return phase. This will mark the completion of this search process.

 \subsection{Back to \eref{s22}}
We first note that the sum of two terms in
  \eref{s6} and \eref{s7} give us
 \begin{align}
     &\langle I(R\le T)I(T_{ret}^O(\Vec{x})\le T_{ret}^A(\Vec{x}))T_{ret}^O(\Vec{x}) + I(R\le T)I(T_{ret}^A(\Vec{x})< T_{ret}^O(\Vec{x}))T_{ret}^A (\Vec{x})\rangle \nonumber \\
     =&\int_{\mathcal{D}}d\Vec{x}\widetilde{G}_R(\Vec{x})\Big[ Pr(T_{ret}^O(\Vec{x})\le T_{ret}^A(\Vec{x}))\langle T_{ret}^O(\Vec{x})|T_{ret}^O(\Vec{x})\le T_{ret}^A(\Vec{x})\rangle& \nonumber \\
         & \hspace{5cm}+ Pr(T_{ret}^A(\Vec{x})< T_{ret}^O(\Vec{x}))\langle T_{ret}^A(\Vec{x})|T_{ret}^A(\Vec{x})< T_{ret}^O(\Vec{x})\rangle\Big] \nonumber \\
         =&\int_{\mathcal{D}}d\Vec{x}\widetilde{G}_R(\Vec{x}) \langle min\left(T_{ret}^A(\Vec{x}),T_{ret}^O(\Vec{x})\right)\rangle,
         \label{s8}
 \end{align}
 where we have used the definition from \eref{min}. For brevity, we have also omitted all the subscripts from the averages. 
 
 Finally, combining together \eref{s5} and \eref{s8} into \eref{s22}, we arrive at the Eq. (4) of the main text
  \begin{align}
\langle T_R \rangle =\frac{\langle min(T,R) \rangle +\int_{\mathcal{D}}d\Vec{x}\widetilde{G}_R(\Vec{x})\langle min\left(T_{ret}^A(\Vec{x}),T_{ret}^O(\Vec{x})\right)\rangle}{1-\int_{\mathcal{D}}d\Vec{x}~\widetilde{G}_R(\Vec{x}) Pr(T_{ret}^O(\Vec{x})\le T_{ret}^A(\Vec{x}))} .
     \label{s10}
\end{align}

\section{Derivation of $\langle min(T,R) \rangle$ for exponential resetting times}\label{S0}\label{S3}
By definition, the expectation of the random variable $ min(T,R) $ can be written as
\begin{align}
      \langle min(T,R) \rangle =\int_0^\infty dt Pr(T\ge t)Pr(R\ge t).
       \label{prz}
\end{align}
In the case of exponential resetting rate, we have $f_R(t)=re^{-rt}$ where $r$ is the resetting rate. Recalling $f_T(t)=-\frac{dPr(T\ge t)}{dt}$ to be the first passage time distribution of the underlying parent process, we obtain from \eref{prz}
\begin{align}
     \langle min(T,R) \rangle &= \int_0^\infty dt e^{-rt} \int_t^\infty dt'f_T(t') \nonumber \\
     &=\frac{1}{r}-\frac{1}{r}\int_0^\infty dt e^{-rt}f_T(t) \nonumber \\
     &=\frac{1-\widetilde{T}(r)}{r},
     \label{mintr}
\end{align}
where $\widetilde{T}(r)=\int_0^\infty dt e^{-rt}f_T(t)$
is the Laplace transform of $f_T(t)$. For the case of 1d diffusion $f_T(t)$ is given by L\'evy Smirnov distribution as follows
\begin{align}
    f_T(t)= \frac{L}{\sqrt{4\pi D t^3}}e^{-L^2/4Dt},
\end{align}
with the following Laplace transform 
\begin{align}
 \widetilde{T}(r)=e^{-\sqrt{rL^2/D}} \label{lsfpt}  .
\end{align}
Combining the above with \eref{mintr} gives the result mentioned in the main text
\begin{align}
    \langle min(T,R) \rangle=\frac{1}{r} \left(1- e^{-\sqrt{rL^2/D}} \right). \label{mintr2}
\end{align}
At this point, we also find $Pr(T<R)$ which will be useful later on. 
\begin{align}
    & Pr(T<R)=\int_0^{\infty}dt f_T(t) Pr(R>t).
\end{align}
For exponential resetting times $Pr(R>t)=e^{-rt}$ and we have
\begin{align}
    Pr(T<R)&=\int_0^{\infty}dt f_T(t) e^{- r t} =\widetilde{T}(r). \label{ptlr}
\end{align}

\section{General expression for MFPT with instantaneous return}
In the case when the searcher returns to the starting position instantaneously, the return time is exactly zero. That implies the second term in the numerator of \eref{s10} is zero i.e. $\langle min\left(T_{ret}^A(\Vec{x}),T_{ret}^O(\Vec{x})\right)\rangle=0$. Alongside that as the return to origin is guaranteed we also have $Pr(T_{ret}^O(\Vec{x})\le T_{ret}^A(\Vec{x}))=1$. Applying these conditions in \eref{s10}, we get
\begin{align}
    \langle T_R^{inst} \rangle =\frac{\langle min(T,R) \rangle }{1-\int_{\mathcal{D}}d\Vec{x}~\widetilde{G}_R(\Vec{x}) }. \label{inst1}
\end{align}
The second term in the denominator can be simplified as the following
\begin{align}
    \int_{\mathcal{D}}d\Vec{x}~\widetilde{G}_R(\Vec{x})&=\int_{\mathcal{D}}d\Vec{x} \int_{0}^{\infty}dt f_R(t) G(\Vec{x},t) \nonumber\\
    &=\int_{0}^{\infty}dt f_R(t) \int_{\mathcal{D}}d\Vec{x} ~G(\Vec{x},t). 
\end{align}
Note that $\int_{\mathcal{D}}d\Vec{x} ~G(\Vec{x},t)$ is nothing but the survival probability $Q(t)$ defined in \eref{surv}. Thus, we have
\begin{align}
    \int_{\mathcal{D}}d\Vec{x}~\widetilde{G}_R(\Vec{x})&=\int_{0}^{\infty}dt f_R(t) Q(t) \nonumber\\
    &=\int_{0}^{\infty}dt f_R(t) Pr(T\ge t) \nonumber\\
    &=Pr(T\ge R) \nonumber\\
    &=1-Pr(T<R).
    \label{intgr}
\end{align}
Substituting the above  into \eref{inst1}, we recover the MFPT for resetting under instantaneous return
\begin{align}
     \langle T_R^{inst} \rangle =\frac{\langle min(T,R) \rangle }{Pr(T<R) }.
     \label{inst}
\end{align}
Further, for exponential waiting time between resetting events, we can use \eref{mintr} and \eref{ptlr} to recover
\begin{align}
     \langle T_R^{inst} \rangle=\frac{1-\widetilde{T}(r)}{r \widetilde{T}(r)}. \label{tinst}
\end{align}

\section{Derivation of the universal criterion (Eq. (5) in the main text)}
In this section, we derive the criterion presented in Eq. (5). The criterion states that for stochastic return to expedite the  instantaneous return one should have
\begin{align}
     &\langle T_R \rangle < \langle T^{inst}_R \rangle.
     \label{eq:criterion}
\end{align}
Rewriting \eref{s10} as 
\begin{align}
    \langle T_R \rangle=\frac{\langle min (T,R) \rangle + \langle min\left(T_{ret}^A,T_{ret}^O\right) \rangle}{1-Pr(T_{ret}^O\le T_{ret}^A)}
    \label{s33},\end{align} 
where we have used the following
\begin{align}
    \langle min\left(T_{ret}^A,T_{ret}^O\right) \rangle=\int_{\mathcal{D}}d\Vec{x}\widetilde{G}_R(\Vec{x})\langle min\left(T_{ret}^A(\Vec{x}),T_{ret}^O(\Vec{x})\right)\rangle,
\end{align}
and 
\begin{align}
Pr(T_{ret}^O\le T_{ret}^A)=\int_{\mathcal{D}}d\Vec{x}~\widetilde{G}_R(\Vec{x}) Pr(T_{ret}^O(\Vec{x})\le T_{ret}^A(\Vec{x})) .
\end{align}
Using \eref{inst} and \eref{s33} in \eref{eq:criterion}, we have
\begin{align}
   \frac{\langle min (T,R) \rangle + \langle min\left(T_{ret}^A,T_{ret}^O\right) \rangle}{1-Pr(T_{ret}^O\le T_{ret}^A)}< \frac{\langle min(T,R) \rangle }{Pr(T<R) }. \label{s35}
\end{align}
Note 
\begin{align}
   Pr(T_{ret}^O\le T_{ret}^A)&=\int_{\mathcal{D}}d\Vec{x} ~\widetilde{G}_R(\Vec{x}) Pr(T_{ret}^O(\Vec{x})\le T_{ret}^A(\Vec{x})) \nonumber\\
   &=\int_{\mathcal{D}}d\Vec{x} ~\widetilde{G}_R(\Vec{x}) [1-Pr(T_{ret}^A(\Vec{x})<T_{ret}^O(\Vec{x}))] \nonumber\\
   &= \int_{\mathcal{D}}d\Vec{x}~\widetilde{G}_R(\Vec{x})-\int_{\mathcal{D}}d\Vec{x} ~\widetilde{G}_R(\Vec{x}) Pr(T_{ret}^A(\Vec{x})<T_{ret}^O(\Vec{x})) \nonumber\\
   &=1-Pr(T<R)-Pr(T_{ret}^A<T_{ret}^O), \hspace{0.5cm} \text{using \eref{intgr}} .
\end{align}
One can now rearrange \eref{s35} to find
\begin{align}
     &\frac{\langle min (T,R) \rangle + \langle min\left(T_{ret}^A,T_{ret}^O\right) \rangle}{Pr(T<R)+Pr(T_{ret}^A<T_{ret}^O)}< \frac{\langle min(T,R) \rangle }{Pr(T<R) } \nonumber\\
     \implies & \mathcal{T} \equiv \frac{
\langle min\left(T_{ret}^A,T_{ret}^O\right) \rangle}{Pr(T_{ret}^A<T_{ret}^O)} <  \frac{\langle min(T,R) \rangle }{Pr(T<R) } = \langle T^{inst}_R \rangle,
\label{final-cri}
\end{align}
which is the general criterion Eq. (5) announced in the main text.

\section{First passage statistics of the diffusing particle during the return phase} \label{sec3}
In the main text, we have examined the paradigm of a 1d diffusive search process (designated by the diffusion constant $D$) in which
a particle starts at the origin $O$ and continues to diffuse until it hits a
stationary target at a location $L$. In addition, we assumed that the process is reset at a constant rate $r$ (i.e., resetting time density $f_R(t)=re^{-rt}$) upon which a potential $U(x)=\lambda |x|$ centered at the origin is turned on. The particle diffuses through the potential and it is switched off when the particle makes a first return to the origin. Subsequently, the particle resumes its diffusive search phase. Notably, during the return phase the particle also has a finite probability to get absorbed at $L$. Thus, the return phase set-up is akin to a diffusing particle in a confining interval. The interval has two boundaries both of which serve as absorbing boundaries. 
Being absorbed at $L$ marks the completion of the full process while absorption at the origin renews a trial where the diffusive search phase restarts. For this set-up, several quantities such as the mean first time to any of the boundaries, conditional times and splitting probabilities to to each of the boundaries were used in the main text. The aim of this section is to derive these quantities from scratch. Some of these results can also be found in \cite{redner2001}.


Let us denote the propagator in the return phase by $G_{ret}(x,t)$ in the presence of two absorbing boundaries at $x=0$ and $x=L$. We start by recalling that the Fokker-Planck equation for the probability distribution function $G_{ret}(x,t)$ which can be written as \cite{redner2001}
\begin{align}
    \frac{\partial G_{ret}(x,t)}{\partial t}-\lambda \frac{\partial G_{ret}(x,t)}{\partial x}=D\frac{\partial^2 G_{ret}(x,t)}{\partial x^2}
    \label{eq29},
\end{align}
with the initial condition 
\begin{align}
    G_{ret}(x,t=0)=\delta (x-x_0), \label{ic}
\end{align} 
where $x_0$ is the position of the particle at the time of resetting. The boundary conditions read
\begin{align}
    G_{ret}(x=0,t)=G_{ret}(x=L,t)=0. \label{bc}
\end{align} 
In Laplace space, \eref{eq29} reads
\begin{align}
    D\frac{\partial^2 \widetilde{G}_{ret}(x,s)}{\partial x^2}+\lambda \frac{\widetilde{G}_{ret}(x,s)}{\partial x}-s\widetilde{G}_{ret}(x,s)=-\delta (x-x_0), \label{GLS}
\end{align}
where $\widetilde{G}_{ret}(x,s)=\int_0^\infty G_{ret}(x,t)e^{-st}dt$ is the Laplace transform of the propagator $G_{ret}(x,t)$. 
We can solve \eref{GLS} by employing the standard Green's function method which gives \begin{align}
   \widetilde{G}_{ret}(x,s)&= -\frac{e^{\frac{\lambda  (x_0-x)}{2 D}} \text{csch}\left(mL\right)}{2mD} \Bigg[\theta (x-x_0) \Big(\cosh \left(m(L+x-x_0)\right)-\cosh \left(m(L-x+x_0)\right)\Big)\nonumber \\
   & \hspace{4cm}+ \cosh \left(m(L-x-x_0)\right)-\cosh \left(m(L+x-x_0)\right)\Bigg],
   \label{gls}
\end{align}
where, $m=\frac{\sqrt{\lambda^2 + 4Ds}}{2D}$ and $\theta(x)$ is the Heaviside step function.

\subsection{The mean first passage time}
The survival probability $Q(t)$ of the particle in this set up is given as
\begin{align}
   Q(t)=\int_{0}^L~dx~G_{ret}(x,t),
\end{align}
which in Laplace space reads
\begin{align}
    \widetilde{Q}(s)=\int_0^{\infty}dt e^{-st} Q(t)=\int_{0}^L \widetilde{G}_{ret}(x,s) dx.
\end{align}
The unconditional mean first passage time that the particle can exit through any of the boundaries is given by
\begin{align}
   \langle t_2(x_0) \rangle&= \lim_{s\to 0}  \widetilde{Q}(s)
   =\frac{L(1-e^{\lambda x_0/D}) +x_0(e^{\lambda L/D }-1)}{\lambda(e^{\lambda L/D}-1)},
   \label{ucmfpt}
\end{align}
which was used in the main text. When the particle is at the negative side quadrant, it only experiences the boundary at the origin. The mean time for this particle to the origin can simply be obtained by setting $L\to \infty$ so that we have
\begin{align}
    \langle t_1(x_0) \rangle=\frac{|x_0|}{\lambda}, \label{ucmfpt1}
\end{align}
which was used in the main text.

\subsection{Splitting probabilities}
The probability flux through each of the boundaries in Laplace space is given by \cite{redner2001}
\begin{align}
    j_L(x_0,s)&=-D\frac{\partial  \widetilde{G}_{ret}(x,s)}{\partial x}\bigg|_{x=L}, \label{jl}\\
        j_O(x_0,t)&=D\frac{\partial  \widetilde{G}_{ret}(x,s)}{\partial x}\bigg|_{x=0}.
        \label{jo}
\end{align}
Using \eref{gls}, (\ref{jl}), (\ref{jo}), one can immediately write the corresponding exit/splitting probabilities through each of the boundaries. These read
\begin{align}
   \epsilon_L(x_0)=j_L(x_0,s\to 0)&=\frac{1-e^{\lambda x_0/D}}{1-e^{\lambda L/D}}, \label{el}\\
    \epsilon_O(x_0)=j_O(x_0,s\to 0)&=1-\epsilon_L(x_0) =\frac{e^{\lambda x_0/D}-e^{\lambda L/D}}{1-e^{\lambda L/D}}. \label{eo}
\end{align}

\section{Derivation of Eq. (6) in the main text}\label{SC2}
In this section, we derive Eq. (6) in the main text for one-dimensional (1d) search process. However, before doing so we make a general discussion on the splitting probabilities and conditional times for a generic stochastic process.

\subsection{Splitting probabilities -- general definition}
For any arbitrary stochastic process in the presence of multiple targets (and not necessarily a diffusive process), a generic definition for the splitting probabilities can be made in the following way
\begin{align}
    \epsilon_A(x)=Pr(T_{ret}^A(x)< T_{ret}^O(x)),
\end{align}
where $T_{ret}^i(x)$ is the random time to reach  $i \in \{A,O\}$, starting from a coordinate $x$. Thus, splitting probability to the target $A$ is conditioned on the fact that $T_{ret}^A(x)< T_{ret}^O(x)$. 
Similarly, 
\begin{align}
\epsilon_O(x)=Pr(T_{ret}^O(x)\le T_{ret}^A(x))
\end{align}
is the splitting probability to reach the target $O$ before it hits the other boundaries. For diffusive process, these were computed in the previous section.

\subsection{Conditional and unconditional mean first-passage/exit times -- general definition}
Similar to the above, one can also consider the average of the conditional or first passage times either to the origin or any of the boundaries. For example, the conditional mean time to reach the origin before it reaches the boundary $A$, starting from $x$, can be written as
\begin{align}
    \langle T_{ret}^O(x)|T_{ret}^O(x)\le T_{ret}^A(x) \rangle= \langle t(x) \rangle_O.
\end{align}
Similarly, 
the conditional time to reach the boundary  $\{ A \}$ before it hits the origin $O$ is given by
\begin{align}
    \langle T_{ret}^A(x)|T_{ret}^A(x)<T_{ret}^O(x) \rangle= \langle t(x) \rangle_A.
\end{align}
Thus, the unconditional mean first passage time to exit through any of the boundaries without any preference, starting from $x$, is given by $\epsilon_A(x) \langle t(x) \rangle_A+ \epsilon_O(x)
    \langle t(x) \rangle_O$.

\subsection{Back to the derivation of Eq. (6)}
Before we derive Eq. (6), let us recall the one-dimensional set-up again from Sec. \ref{sec3} and in particular, note that the target ($A$) is located at $L>0$. Thus, if the searcher starts from $x<0$ in the return phase, it will not find $L$ and thus 
$Pr(T_{ret}^O(x)\le T_{ret}^L(x))=1$ \text{~and~} $Pr(T_{ret}^L(x)< T_{ret}^O(x))=0$ so that
\begin{align}
\langle min\left(T_{ret}^L(x),T_{ret}^O(x)\right)\rangle=\langle T_{ret}^O(x) \rangle_{return}= \langle t_1(x) \rangle, \hspace{1cm} \text{for } x<0,
\end{align}
where  $\langle t_1(x) \rangle$, as defined before, denotes the MFPT of the searcher with only one virtual absorbing boundary at the origin $O$ during the return phase. 


If the searcher starts from $x>0$, the situation is subtle since it has both the possibilities of reaching the target ($L$) or the origin ($O$) during the return. 
In this case, 
\begin{align}
&\langle min\left(T_{ret}^L(x),T_{ret}^O(x)\right)\rangle\nonumber \\
    =&~Pr(T_{ret}^O(x)\le T_{ret}^L(x)) \langle T_{ret}^O(x)|T_{ret}^O(x)\le T_{ret}^L(x) \rangle +Pr(T_{ret}^L(x)< T_{ret}^O(x)) \langle T_{ret}^L(x)|T_{ret}^L(x)< T_{ret}^O(x) \rangle\nonumber \\
    =&~ \epsilon_O(x) \langle t(x) \rangle_O + \epsilon_L(x) \langle t(x) \rangle_L \nonumber \\
    =&~\langle t_2(x) \rangle, \hspace{1cm} \text{for } x>0,
\end{align}
where $\langle t_2(x) \rangle$, as defined earlier, denotes the unconditional MFPT of a searcher to exit through any of the two boundaries at $x=0$ and $x=L$ during the return phase.
Combining the results for $x<0$ and $x>0$, we can write the second term in the numerator of \eref{s10} as
\begin{align}
   \langle min\left(T_{ret}^L(x),T_{ret}^O(x)\right)\rangle=\theta(-x)\langle t_1(x) \rangle + \theta(x)\langle t_2(x) \rangle ,
    \label{mintlto}
\end{align}
which is Eq. (6) in the main text. Similarly, the second term in the denominator of Eq. (4) (or \eref{s10}) can be written as
\begin{align}
    Pr(T_{ret}^O(x)\le T_{ret}^L(x))=\theta (-x)+ \theta(x) \epsilon_O(x).
    \label{prtlto}
\end{align}
\eref{mintlto} and \eref{prtlto} combindedly give the one-dimensional form of the MFPT in \eref{s10} as 
\begin{align}
    \langle T_R \rangle =\frac{\langle min(T,R) \rangle +\int_{-\infty}^Ldx\widetilde{G}_R(x)\left[\theta(-x)\langle t_1(x) \rangle + \theta(x)\langle t_2(x) \rangle\right]}{1-\int_{-\infty}^Ldx~\widetilde{G}_R(x) \left[\theta (-x)+ \theta(x) \epsilon_O(x)\right]}.
    \label{s24}
\end{align}
In case of exponential resetting times $\widetilde{G}_R(x)=r \widetilde{G}(x,r)$ where
\begin{align}
    \widetilde{G}(x,r)=\int_0^\infty~dt~e^{-rt}~G(x,t)
\end{align}
is the Laplace transform of $G(x,t)$ -- the propagator for the reset free process. Eq. (\ref{s24}) was used to derive Eq. (7) in the main text.

\section{Variation of MFPT $\langle \tau (\overline{r}, \overline{\lambda}) \rangle$ with  resetting rate $\overline{r}$}
In this section, we show how MFPT $\langle \tau (\overline{r}, \overline{\lambda}) \rangle$ as in Eq. (6) of the main text, varies with respect to the resetting rate $\overline{r}=\frac{r L^2}{D}$ for different values of $\overline{\lambda}=\frac{\lambda L}{D}$.
\begin{figure}[H]
    \centering
    \includegraphics[width=9cm]{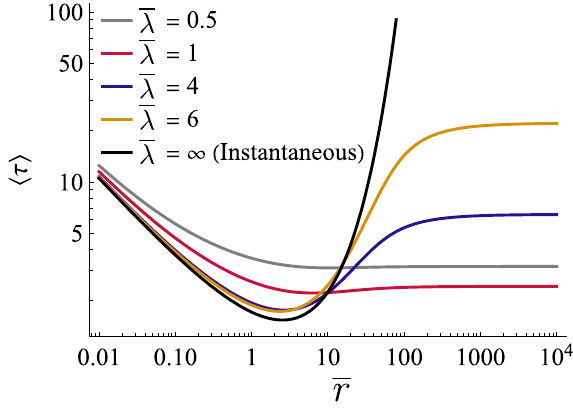}
    \caption{ Behaviour of  MFPT $\langle \tau (\overline{r}, \overline{\lambda}) \rangle$ with respect to $\overline{r}$ for different values of $\overline{\lambda}$. The black line corresponds to $\overline{\lambda}\to \infty$ which is the classical Evan-Majumdar result for instantaneous return \cite{evans_diffusion_2011}. }
    \label{figs1}
\end{figure}
Let us first discuss the limiting cases. For low enough values of $\overline{r}$ the MFPT diverges as the particle can disperse much distance away from the target. For very large values of $\overline{r}$ the potential is effectively  always turned on and the corresponding MFPT saturates to a finite value given by (also shown in main text)
\begin{align}
     \langle \tau(\overline{r}\to\infty,\overline{\lambda})\rangle=\frac{2 e^{\overline{\lambda}}-\overline{\lambda}-2}{\overline{\lambda}^2}.
\end{align}
The MFPT shows a non-monotonic behaviour with respect to $\overline{\lambda}$. For low $\overline{\lambda}$ , the system is effectively diffusive and thus, in the absence of resetting, MFPT is infinite. Even with resetting (and low $\overline{\lambda}$ number), the system effectively remains diffusive since the particle experiences a very weak attraction to the origin. In other words, both the search and return phase remain diffusive resulting in a high MFPT. For very high $\overline{\lambda}$, the return is instantaneous and we recover the result of the classical first passage under resetting \cite{evans_diffusion_2011}. 




\section{MFPT for resetting with deterministic return in 1d}
For the deterministic return i.e., when the particle returns to the origin following some deterministic dynamics, there is zero probability of target finding resulting in
\begin{align}
    \epsilon_L(x)=1-\epsilon_O(x)=0.
\end{align}
Henceforth, for the diffusing Brownian particle that returns to origin with a constant velocity $\lambda$ from $x$, the return time is given by
\begin{align}
    \langle t_1(x)\rangle=\langle t_2(x)\rangle=\frac{|x|}{\lambda}.
\end{align}
Note that unlike stochastic return, cases for the $x>0$ and $x<0$ for the deterministic return are identical. Substituting the above expressions into \eref{mintr} and \eref{s24} we obtain the MFPT as
\begin{align}
     \langle T_{det}\rangle =  \frac{\frac{1-\widetilde{T}(r)}{r}  +r\int_{{-\infty}}^L dx \widetilde{G}(x,r)\frac{|x|}{\lambda}}{1-\int_{-\infty}^Ldx~\widetilde{G}(x,r)},\label{s27}
\end{align}
where $\widetilde{G}(x,r)$ and $\widetilde{T}(r)$ are the Laplace transform of the reset free process propagator and the first passage time distribution respectively. Here invoking \eref{intgr} and \eref{ptlr} we can replace the denominator with simply $\widetilde{T}(r)$ to have
\begin{align}
     \langle T_{det}\rangle =  \frac{\frac{1-\widetilde{T}(r)}{r}  +r\int_{{-\infty}}^L dx \widetilde{G}(x,r)\frac{|x|}{\lambda}}{\widetilde{T}(r)} \label{tdet}.
\end{align}
For the diffusive dynamics with propagator
\begin{align}
    G(x,t)&=\frac{1}{\sqrt{4\pi Dt}}\left(e^{-\frac{x^2}{4Dt}} -e^{-\frac{(2L-x)^2}{4Dt}}\right), \nonumber\\
   \implies \widetilde{G}(x,r)&=\frac{1}{\sqrt{4Dr}}\left(e^{-\sqrt{r/D}|x|}-e^{-\sqrt{r/D}(2L-x)}\right), \label{frp}
\end{align}
and \eref{lsfpt}, we can solve the intergrals in \eref{tdet} to arrive at the following expression for MFPT with deterministic return in dimensionless form
\begin{align}
    \langle \tau_{det}(\overline{r},\overline{\lambda}) \rangle=\frac{D \langle T_{det} \rangle }{L^2}=\frac{e^{\sqrt{\overline{r}}}-1}{\overline{r}}+\frac{1}{\overline{\lambda}} \left(\frac{2 \sinh \left(\sqrt{\overline{r}}\right)}{\sqrt{\overline{r}}}-1 \right),
    \label{mfptdet}
    \end{align}
where recall $\overline{r}=\frac{rL^2}{D}$ and $\overline{\lambda}=\frac{\lambda L}{D}$ (also obtained in \cite{pal_search_2020}).

\section{Limiting behaviours of $\langle \tau(\overline{r},\overline{\lambda}) \rangle$ with respect to $\overline{\lambda}$ for different $\overline{r}$}

\begin{figure}[H]
    \centering
    \includegraphics[width=16cm]{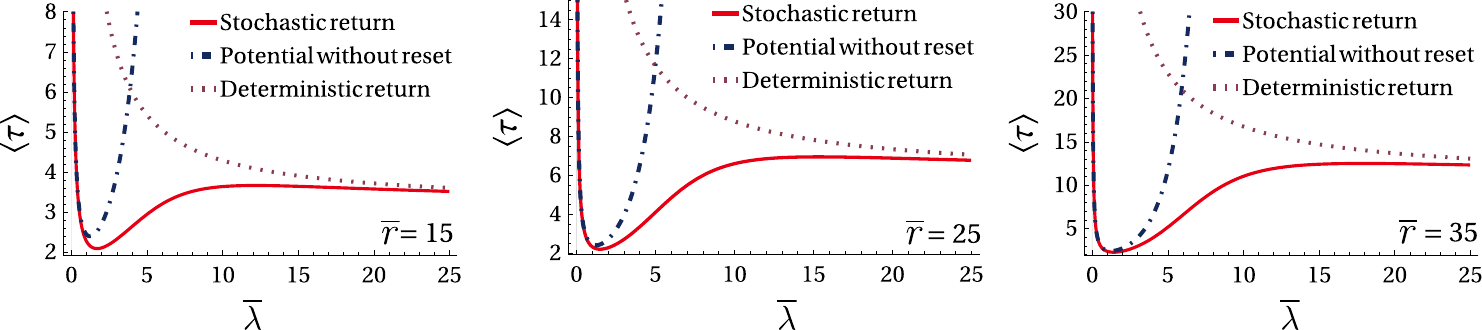}
    \caption{A systematic comparison between the MFPT as a function of $\overline{\lambda}$ for three different resetting rates. This plot extends Fig. 2 in the main text where we have shown how the shape of the MFPT can be understood from two distinct physical limits namely the MFPT with deterministic returns and the MFPT under potential without resetting. The right and left envelopes correspond to each of these limits respectively. Finally, as the resetting rate $\overline{r}$ is increased the overlapping becomes more prominent.}
    \label{comp}
\end{figure}

In this section, we illustrate further detail on the limiting behaviour of MFPT of stochastic return for the 1d diffusive system with respect to the $\overline{\lambda}$ number. In particular, several comments can be made on the shape of the MFPT curve in Fig. 2 of the main text. To illustrate further, we reproduce similar plots for different values of $\overline{r}$ in  Fig. (\ref{comp}). Each panel in Fig. (\ref{comp}) shows variation of $\langle \tau(\overline{r},\overline{\lambda}) \rangle$, obtained from Eq. (7) in the main text  (solid red line), as a function of the $\overline{\lambda}$ number.

In the limit of large $\overline{\lambda}$ (i.e., for the strongly attractive trap), the particle returns to the origin almost deterministically with constant speed $\lambda$ and has negligible probability to find the target during the return phase. Therefore, one can approximate the right envelope with $ \langle \tau_{det}(\overline{r},\overline{\lambda}) \rangle=\frac{e^{\sqrt{\overline{r}}}-1}{\overline{r}}+\frac{1}{\overline{\lambda}} \left(\frac{2 \sinh \left(\sqrt{\overline{r}}\right)}{\sqrt{\overline{r}}}-1 \right)$, as derived in Eq. (\ref{mfptdet}). This is shown by the dotted line in each panel in Fig. (\ref{comp}).

On the other hand, for smaller $\overline{\lambda}$ values, the potential is rather shallow and it takes time for the particle either to return to the origin or to the target. However, if $\overline{r}$ is large, the resetting events are more frequent and thus the contribution to the global MFPT comes predominantly from the post-resetting phase which can be computed from Eq. (7) as $ \langle \tau(\overline{r}\to\infty,\overline{\lambda})\rangle=\frac{2 e^{\overline{\lambda}}-\overline{\lambda}-2}{\overline{\lambda}^2}$. This is shown by the dot-dashed line in each panel in Fig. (\ref{comp}). The agreement is found to be excellent between the two.


Note that the curve for MFPT (with potential always ON) overlaps (the left envelope of the solid curve) more and more to the MFPT under stochastic return as one increases the resetting rate $\overline{r}$ (going from the left panel to the right). When $\overline{\lambda}$ is sufficiently strong there is almost a negligible probability for the target detection during the return phase and the particle reaches the origin with constant speed (due to the linear nature of the potential). Thus, this limit (i.e, the right envelope of the solid curve) resonates with the deterministic return problem \cite{pal_search_2020}.

\section{Application of the criterion in Eq. (5) for diffusive search}
In this section, we illustrate the criterion derived in the previous section for the 1d diffusive search. To this end, we compute each term in \eqref{final-cri} which is also Eq. (5) in the main text. For instance,
\begin{align}
    \mathcal{T}=\frac{
\langle min\left(T_{ret}^L,T_{ret}^O\right) \rangle}{Pr(T_{ret}^L<T_{ret}^O)} =\frac{\int_{-\infty}^L dx \widetilde{G}(x,r)\Big[\theta(-x)\langle t_1(x) \rangle + \theta(x)\langle t_2(x) \rangle \Big]}{\int_{0}^{L} dx~\widetilde{G}(x,r) \epsilon_L(x)},
    \label{sr1d}
\end{align}
where we have used \eref{mintlto} and 
\begin{align}
Pr(T_{ret}^L<T_{ret}^O)&=\int_{-\infty}^Ldx~\widetilde{G}(x,r) Pr(T_{ret}^L(x)<T_{ret}^O(x)) \nonumber\\
&=\int_{0}^{L} dx~\widetilde{G}(x,r) \epsilon_L(x).
\end{align}
Note the negative part of the integration vanishes as the probability of reaching $L$ is simply zero there during the return phase.

One can now plugin the propagator for diffusion (in Laplace space) i.e., $\widetilde{G}(x,r)$ from \eref{frp}, the unconditional MFPTs $\langle t_1(x) \rangle, \langle t_2(x) \rangle$ from \eref{ucmfpt1}, \eref{ucmfpt}, the splitting probability to target at $L$ i.e. $\epsilon_L(x)$ from \eref{el} and $\widetilde{T}(r)$ from \eref{lsfpt}. The quantity $\mathcal{T}$ then takes the dimensionless form (scaled by $L^2/D$)\begin{small}
\begin{align}
   \mathcal{T}(\overline{r},\overline{\lambda})= \frac{\overline{\lambda}^2 \left(2 \sqrt{\overline{r}} e^{\overline{\lambda}+\sqrt{\overline{r}}}-e^{2 \sqrt{\overline{r}}} \left(2 e^{\overline{\lambda}}+\sqrt{\overline{r}}-2\right)+2 e^{\overline{\lambda}}-\sqrt{\overline{r}}-2\right)-\overline{\lambda} \left(e^{2 \sqrt{\overline{r}}}-1\right) \overline{r}+2 \left(e^{\overline{\lambda}}-1\right) \left(e^{2 \sqrt{\overline{r}}}-1\right) \overline{r}}{\overline{\lambda} \sqrt{\overline{r}} \left(\overline{\lambda}^2 \left(e^{\sqrt{\overline{r}}}-1\right)^2+\overline{\lambda} \left(e^{2 \sqrt{\overline{r}}}-1\right) \sqrt{\overline{r}}-2 \left(e^{\overline{\lambda}}-1\right) e^{\sqrt{\overline{r}}} \overline{r}\right)},
\end{align}
\end{small}whereas the RHS which is the MFPT of the instantaneous return can be found by inserting \eref{lsfpt} in \eref{tinst} (or taking the limit $\overline{\lambda}\to\infty$ in Eq. (6) of  main text)  as
\begin{align}
   \langle\tau_{inst}(\overline{r}) \rangle=\frac{D}{L^2}\langle T^{inst}_R \rangle =\frac{e^{\sqrt{\overline{r}}}-1}{\overline{r}}.
\end{align}
The criterion in Eq. (5) (or \eref{final-cri}) holds when $\frac{\mathcal{T}(\overline{r},\overline{\lambda})}{\langle\tau_{inst}(\overline{r}) \rangle}<1$ and stochastic return facilitates first passage over instantaneous return. The equality 
\begin{align}
\mathcal{T}(\overline{r},\overline{\lambda})= \langle\tau_{inst}(\overline{r}) \rangle,    
\end{align}
gives a solution for $\overline{r}^*$ for each $\overline{\lambda}$. Thus, one can generate a separatrix (the red dashed line in Fig. 3a of main text which is essentially the line $\overline{r}^*(\overline{\lambda})$) that distinguishes between two regions -- either stochastic return (SR) or instantaneous return (IR) better. In Table 1,  we show different limits of this ratio and conclude where stochastic return is better than instantaneous return for the diffusive case.
\renewcommand{\arraystretch}{3}
\begin{table}[H]
\centering
\caption*{\textbf{Table 1:}  Demonstration  of criterion in Eq. (5) for 1d diffusion in different limits of $\overline{r}$ and $\overline{\lambda}$}
\begin{adjustbox}{width=15cm}
\begin{tabular}{||c|c|c|c|c||}
\hline
Quantity &$\overline{r}\to 0$, Finite $\overline{\lambda}$ & $\overline{r}\to \infty$, Finite $\overline{\lambda}$ & $\overline{\lambda}\to 0$, Finite $\overline{r}$ &  $\overline{\lambda}\to \infty$, Finite $\overline{r}$ \\
\hline
$\mathcal{T}(\overline{r},\overline{\lambda})$ & $\frac{1}{\overline{r}}\left(\frac{2 \overline{\lambda} \left(1-e^{\overline{\lambda}}\right)  }{\overline{\lambda} (\overline{\lambda}+2)-2 e^{\overline{\lambda}}+2  }\right)$ & $\frac{2 e^{\overline{\lambda}}-\overline{\lambda}-2}{\overline{\lambda}^2}$ & $\frac{1}{\overline{\lambda}}\frac{1}{1-\sqrt{\overline{r}} \text{csch}\left(\sqrt{\overline{r}}\right)}$ & $\overline{\lambda}\left(\frac{2  \sinh \left(\sqrt{\overline{r}}\right)}{\overline{r}^{3/2}}-\frac{1}{\overline{r}}\right)$ \\
\hline
$\langle\tau_{inst}(\overline{r}) \rangle$ & $ \frac{1}{\sqrt{\overline{r}}}$ & $\frac{e^{\sqrt{\overline{r}}}}{\overline{r}}$ & $\frac{e^{\sqrt{\overline{r}}}-1}{\overline{r}}$ & $\frac{e^{\sqrt{\overline{r}}}-1}{\overline{r}}$  \\
\hline
$\frac{\mathcal{T}(\overline{r},\overline{\lambda})}{\langle\tau_{inst}(\overline{r}) \rangle}$ & $>1$, IR better & $<1$, SR better & $>1$, IR better & $>1$, IR better\\
\hline
\end{tabular}
\end{adjustbox}
\end{table} 

\section{First passage time under resetting and stochastic return facilitated with harmonic potential $U(x)=\frac{1}{2}\lambda x^2$}
In the previous sections, we have discussed the diffusive search with resetting and stochastic return under linear potential. Here, we extend our theoretical formalism to incorporate the stochastic return process facilitated by the harmonic potential. Namely, the particle performs diffusion in the search/exploration phase. Following a reset, a harmonic trap (centered at the origin) is switched on and this marks the inception of the return phase. Thus, the return process is, in effect, an Ornstein-Uhlenbeck (OU) process. As mentioned earlier, during the return phase the particle has two possibilities: either to return to the origin and resume its search or to capture the target and end the search. Besides being a paradigm of non-equilibrium process, the OU process holds a special place in physics since the thermal position fluctuations of a colloidal particle in an experimentally calibrated optical trap can essentially be described by the OU process.

The aims of this section are two folds: first, is to derive the total mean first passage time under resetting using the formalism developed in the main text; second, is to obtain a phase diagram for search efficiency for a realistic set of parameters that were used in recent resetting experiments \cite{besga2020optimal,faisant2021optimal}.  

\subsection{Analytical derivation of the MFPT}
\begin{figure}[H]
    \centering
    \includegraphics[width=9cm]{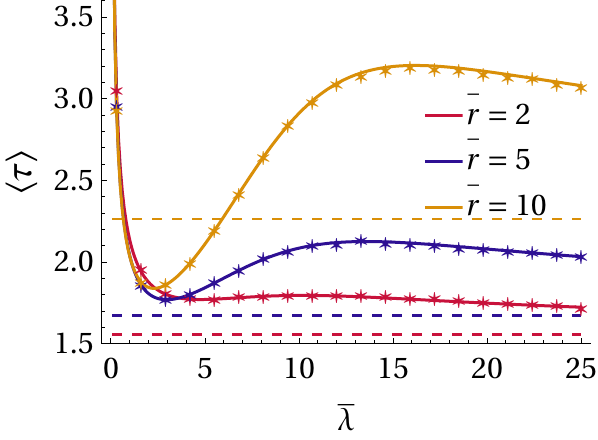}
    \caption{Variation of MFPT $\langle \tau(\overline{r},\overline{\lambda}) \rangle$ for the harmonic case as in Eq. (\ref{MFPT-harmonic}) as a function of the potential strength $\overline{\lambda}=\lambda L^2/D$, for different values of resetting rate $\overline{r}=rL^2/D$. The dashed horizontal lines represent the corresponding MFPT in the limit of instantaneous return for each resetting rate. Note that the plot is qualitatively similar to Fig 2 of the main text which was for the linear potential. The markers are from numerical simulations and the solid lines are obtained from theory. We find an excellent match between theory and simulation.}
    \label{figs3}
\end{figure}

To derive the MFPT in the case of harmonic trap, we follow the exact steps that were used in Sec \ref{sec3}. We start by computing the propagator $G_{ret}(x,t)$ in the return phase under harmonic potential. The governing equation for $G_{ret}(x,t)$, similar to \eref{eq29}), reads as
\begin{align}
    \frac{\partial G_{ret}(x,t)}{\partial t}- \frac{\partial}{\partial x} \left[ \lambda x G_{ret}(x,t)\right]=D\frac{\partial^2 G_{ret}(x,t)}{\partial x^2},
    \label{shm_prop}
\end{align}
with the same initial condition as in \eref{ic} and boundary conditions as in \eref{bc}. One can then proceed to follow the similar steps as in Sec \ref{sec3} to find both the splitting probabilities and the associated mean first passage times. It 
will prove convenient to represent the observables in terms of the following dimensionless variables
\begin{align}
    \overline{\lambda}=\frac{\lambda L^2}{D},~~ \overline{r}=\frac{r L^2}{D}, ~~ \overline{x}=x/L .
\end{align}
Following Sec \ref{sec3} and skipping the details, we find the 
splitting probabilities to be
\begin{align}
      \epsilon_L(\overline{x})=\frac{\text{erfi}\left(\frac{\overline{x} \sqrt{ \overline{\lambda}}}{\sqrt{2}}\right)}{\text{erfi}\left(\frac{\sqrt{ \overline{\lambda}}}{\sqrt{2}}\right)}, ~~
    \epsilon_O(\overline{x})=1-\frac{\text{erfi}\left(\frac{\overline{x} \sqrt{ \overline{\lambda}}}{\sqrt{2}}\right)}{\text{erfi}\left(\frac{ \sqrt{ \overline{\lambda}}}{\sqrt{2}}\right)}.
    \end{align}
The unconditional mean first passage times for this case can also be computed in a similar manner and they read  
\begin{align}
     \langle t_1(x) \rangle &=\frac{L^2}{D}\mathcal{G}_1(\overline{\lambda},\overline{x}),~~
     \langle t_2(x) \rangle =\frac{L^2}{D}\mathcal{G}_2(\overline{\lambda},\overline{x}),
    \end{align}
where
\begin{align}
 &\mathcal{G}_1(\overline{\lambda},\overline{x})= 
 \frac{\overline{\lambda} \overline{x}^2 \, _2F_2\left(1,1;\frac{3}{2},2;-\frac{1}{2} \left(\overline{x}^2 \overline{\lambda} \right)\right)-\pi  \left(\text{erf}\left(\frac{\sqrt{\overline{\lambda} }\overline{x}}{\sqrt{2}}\right)-1\right) \text{erfi}\left(\frac{\sqrt{\overline{\lambda} }\overline{x}}{\sqrt{2}}\right)}{2 \overline{\lambda} },\\
    &\mathcal{G}_2(\overline{\lambda},\overline{x})=\frac{1}{2} \Bigg(-\frac{\text{erfi}\left(\frac{\sqrt{\overline{\lambda} } \overline{x}}{\sqrt{2}}\right) \, _2F_2\left(1,1;\frac{3}{2},2;-\frac{\overline{\lambda} }{2}\right)}{\text{erfi}\left(\frac{\sqrt{\overline{\lambda} }}{\sqrt{2}}\right)}+\overline{x}^2 \, _2F_2\left(1,1;\frac{3}{2},2;-\frac{1}{2} \left(\overline{x}^2 \overline{\lambda} \right)\right)\nonumber\\
    &\hspace{7cm}+\frac{\pi }{\overline{\lambda}} \left(\text{erf}\left(\frac{\sqrt{\overline{\lambda} }}{\sqrt{2}}\right)-\text{erf}\left(\frac{\sqrt{\overline{\lambda} } \overline{x}}{\sqrt{2}}\right)\right) \text{erfi}\left(\frac{\sqrt{\overline{\lambda} } \overline{x}}{\sqrt{2}}\right)\Bigg),
\end{align}
and $\, _2F_2\left(a,b;c,d;z\right)$ is the generalized hypergeometric function. At this point, we also recall the underlying free range propagator in Laplace space ($\widetilde{G}(x,r)$) from \eref{frp} 
\begin{align}
    \widetilde{G}(x,r)&=\frac{L}{D}\frac{1}{\sqrt{4\overline{r}}}\left(e^{-\sqrt{\overline{r}}|\overline{x}|}-e^{-\sqrt{\overline{r}}(2-\overline{x})}\right)= \frac{L}{D}\widetilde{G}(\overline{x},\overline{r}),
\end{align}
which we have rewritten in terms of the dimensionless parameter $\overline{r}$. Finally, the first term in the numerator of \eref{s24} can be rewritten as
\begin{align}
     \langle min(T,R) \rangle=\frac{L^2}{D} \frac{1}{\overline{r}} \left(1- e^{-\sqrt{\overline{r}}} \right).
\end{align}
With these quantities in hand one can now directly use \eref{s24} to find the total mean first passage time 
\begin{align}
    \langle \tau(\overline{r},\overline{\lambda}) \rangle &=\frac{D}{L^2}\langle T_R \rangle=\frac{ \frac{1}{\overline{r}} \left(1- e^{-\sqrt{\overline{r}}} \right) +\overline{r}\int_{-\infty}^1 d\overline{x} \widetilde{G}(\overline{x},\overline{r})\left[\theta(-\overline{x})\mathcal{G}_1(\overline{\lambda},\overline{x}) + \theta(\overline{x})\mathcal{G}_2(\overline{\lambda},\overline{x})\right]}{1-\overline{r}\int_{-\infty}^1d\overline{x}~ \widetilde{G}(\overline{x},\overline{r}) \left[\theta (-\overline{x})+ \theta(\overline{x}) \epsilon_O(\overline{x})\right]},
    \label{MFPT-harmonic}
\end{align}
where the integrals can be computed numerically. In Fig. \ref{figs3} we show the behaviour of the MFPT $\langle \tau(\overline{r},\overline{\lambda}) \rangle$ with respect to $\overline{\lambda}$ for various value of $\overline{r}$ (similar to Fig. 2 of the main text) and corroborate the results with simulation. Similar to the linear potential case, here too we observe that there is a range of $\overline{\lambda}$ for which the MFPT goes below the yellow dashed line associated to the MFPTs with instantaneous resetting (see the yellow curve for $\overline{r}=10$). Naturally, in this regime stochastic returns turn out to be a better strategy.


\subsection{The phase diagram for search efficiency: identification of realistic range of parameters}
\begin{figure}[H]
    \centering
    \includegraphics[width=9cm]{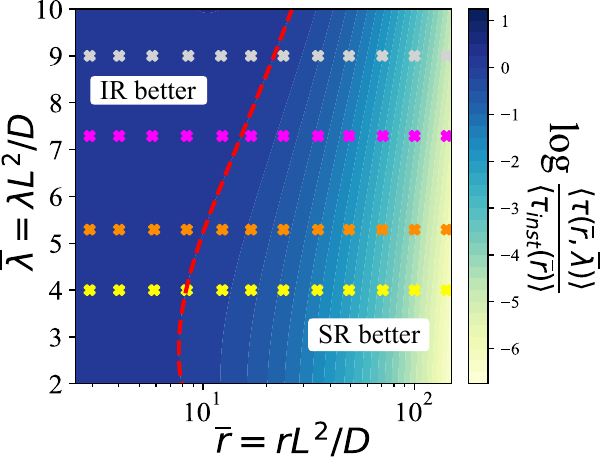}
    \caption{\textbf{The phase diagram} for a diffusive search with stochastic returns facilitated by the harmonic potential. Here, the mean search time $\langle \tau(\overline{r},\overline{\lambda}) \rangle$ can be computed exactly and it is represented in terms of the (i) dimensionless  potential strength $\overline{\lambda}= \lambda L^2/D$ and (ii) the dimensionless resetting rate  $\overline{r}=r L^2/D$. To generate the phase diagram, we utilize the universal criterion Eq 5 from the main text, and vary (i) dimensionless  potential strength $\overline{\lambda}= \lambda  L^2/D$ and (ii) the dimensionless resetting rate  $\overline{r}=r L^2/D$. The red dashed line is the separatrix that distinguishes between two regimes: stochastic return (SR) detrimental over instantaneous return (IR) and stochastic return (SR) beneficial over instantaneous return (IR). The phase space plot is superimposed by the cross markers which are experimental data sets extracted from Refs \cite{besga2020optimal,faisant2021optimal} (see in particular Fig. 5 from Ref \cite{besga2020optimal} and Fig. 6 from Ref \cite{faisant2021optimal}). The markers with colors yellow, orange, magenta and gray respectively correspond to different values of experimentally implemented resetting rate for fixed potential strengths set at  $\overline{\lambda}=4,5.9,7.29,9$ (see Fig. 5 of Ref \cite{besga2020optimal}). }
    \label{figs4}
\end{figure}

We can now employ the criterion i.e., Eq. 5 from the main text to generate the phase diagram for this dynamics. In here, we vary the resetting rate and the potential strength to generate the phase diagram. Such a phase diagram, spanned by the system parameters, allows us to identify the parameter regimes where the stochastic
return can be more useful over the classical instantaneous return with regard to an efficient search. Note however that in this case, we have not arbitrarily varied these parameters, rather we have used a list of the parameters with the exact values that were used in the experiments \cite{besga2020optimal,faisant2021optimal}. Graphical illustration of the phase diagram is made in Fig. \ref{figs4}. \\


\noindent
\textbf{The system set-up and relevant parameters:~}
In the following, we briefly revisit the experimental set-up by Besga \textit{et al} \cite{besga2020optimal,faisant2021optimal}. This will be useful to interlace the resetting experiment with our theoretical modeling. In the experiment, the authors considered an overdamped Brownian particle fluctuating inside a harmonic trap with potential $U(x)=\frac{1}{2}\lambda_e x^2$, 
where $\Gamma$ is the damping constant and the $\lambda_e$ stands for the potential strength as obtained from an optical tweezer by controlling its power. $\zeta(t)$ is a Delta correlated white noise with strength $D$. Here, $D=k_B T_B/\Gamma$ is the diffusion constant where $k_B$ is the Boltzmann constant and  $T_B $ is the ambient temperature. The value of $\Gamma$ can be found from the Stoke's law as $\Gamma=6\pi\eta r$, where $r \approx 1\mu m$ is the radius of the particle (as in the experiment) and $\eta$ is the viscosity of water. Given long enough time the particle equilibrates inside the harmonic trap and its position gets distributed according to the Gibbs–Boltzmann distribution $P_{eq}(x)=\frac{e^{-x^2/2\sigma^2}}{\sqrt{2 \pi \sigma^2}}$, where $\sigma=\sqrt{k_B T_B/\lambda_e}$. The authors design an ``instantaneous resetting'' protocol using the harmonic trap that brings the particle to a fluctuating location (close to the origin) distributed according to the $P_{eq}(x)$ \cite{besga2020optimal,faisant2021optimal}. We do not delve deeper into the experimental details here besides stressing the fact that the resetting there was conducted by the harmonic trap. Moreover, resetting was made ``instantaneous'' by throwing away the so-called pulling/equilibration time while in our case, we assume that the return time can not be thrown away (generically this should be true for any arbitrary stochastic process) and thus by its inclusion we have unveiled new physics as illustrated in our letter. This is a notable distinction between our work and majority of the existing works. 

To summarize this part, our return protocol is facilitated by the same harmonic trap as proposed by the experiments (and thus we use their experimentally accessible parameters); however, the return is not instantaneous but a finite-time process. In the following, we demonstrate how the theoretical and experimental variables are related to each other.  \\

\noindent
\textbf{The dimensionless potential strength:~}
In the experiments \cite{besga2020optimal,faisant2021optimal}, the authors introduced
\begin{align}
    b=L/\sigma,
\end{align}
where $L$ is the distance to the target and recall that $\sigma=\sqrt{k_B T_B/\lambda_e}$ which is also the equilibrium length scale of the particle under the harmonic trap with strength $\lambda_e$. This can be connected to the dimensionless potential strength $ \overline{\lambda}$ that we have defined in the theoretical modeling such as 
\begin{align}
    \overline{\lambda}=\lambda L^2/D=\frac{\lambda_e L^2}{D \Gamma}= \frac{k_B T_B}{D \Gamma }b^2= b^2,
    \label{pot-corr}
\end{align}
where $D\Gamma =k_BT_B$ and the potential strength $\lambda$, in our theoretical model, is related to the same from the experiment as $\lambda=\lambda_e/\Gamma$. This dimensionless relation is quite useful as it allows us to scale $D$ or $L$ without specifying their exact values. In the experiment, $b$ is varied in between $2$ to $3$ which corresponds to an approximated range of $\overline{\lambda}\in (4, 9)$. \\

\noindent
\textbf{The dimensionless resetting rate:~} 
 In a similar vein, the authors in \cite{besga2020optimal,faisant2021optimal} constructed a dimensionless resetting rate in the following way  
\begin{align}
    c=\frac{\sqrt{r} L}{\sqrt{D}},
\end{align}
which upon taking squares on both the sides yield the non-dimensional resetting rate $\overline{r}$ for our theory i.e.,
\begin{align}
    \overline{r}=r L^2/D=c^2.
    \label{rate-corr}
\end{align}
In their experiment, $c$ was varied roughly between $0$ and $14$ which corresponds to the following range for our resetting rate: $\overline{r}\in (0,196)$.\\

\noindent
\textbf{Data extraction and the phase diagram:~}
We have used Fig. 5 from Ref \cite{besga2020optimal} and Fig. 6 from Ref \cite{faisant2021optimal} to extract the parameters $(b,c)$. Using the mapping shown by Eqs. \ref{pot-corr} and \ref{rate-corr}, we have generated a corresponding list of parameters $(\overline{\lambda},\overline{r})$. We have used the parameters from this list to generate the phase diagram in Fig. \ref{figs4} (also Fig 3b in the main text). The red dashed line is the separatrix (generated from Eq. 5 in the main text) that distinguishes between two regimes: (i) left regime where the instantaneous returns are more useful, (ii) right regime where stochastic returns (SR) are found to be more beneficial than the instantaneous return (IR). For each coordinate $(\overline{\lambda},\overline{r})$ in the phase diagram, we compute the log ratio between  $\langle \tau(\overline{r},\overline{\lambda}) \rangle$ and $\langle \tau_{inst}(\overline{r}) \rangle=\langle \tau(\overline{r},\overline{\lambda} \to \infty) \rangle$. Cumulatively, this is shown using colorbars in Fig. \ref{figs4} and Fig. 3b from the main text. It is evident from the figure that in the stochastic return dominated regime, there is a significant improvement in the search efficiency as the mean search times can differ by the order of magnitude. 

This phase space plot is effective in two folds: (i) to identify the parameter range where one needs to design stochastic return protocols, and (ii) to quantify the corresponding speed-up in the search time for each parameter coordinate in the phase space especially pertaining to (i). Thus, this phase diagram can serve as a useful cue for the experimentalists to selectively choose the parameters.

 \end{document}